\documentclass[twocolumn,trackchanges]{aastex702}
\usepackage{amsmath}  % For advanced math symbols, if needed
\usepackage{orcidlink}   % For orcids
\usepackage{booktabs}  %% For tables
\usepackage{CJK}      % for Chinese, Japanese, Korean characters

\usepackage{graphicx}	% Including figure files
\usepackage{amsmath}	% Advanced maths commands
\usepackage{soul}
\usepackage{multirow}
\usepackage[version=4]{mhchem}
\usepackage{hyperref}
\usepackage{amsmath}  % For advanced math symbols, if needed
\usepackage{makecell}

\begin{document}
\begin{CJK*}{UTF8}{gbsn}

%%%%%%%%%%%%%%%%%%% TITLE PAGE %%%%%%%%%%%%%%%%%%%

% Title of the paper, and the short title used in the headers.
\title{JWST MEP - A World Under Spotty Starlight: Detection of  \ce{CO2} and \ce{H2O} in the \\ Hot Saturn WASP-52b with JWST NIRSpec G395H}

\shortauthors{Pan et al.}
\shorttitle{WASP-52b with JWST NIRSpec G395H}

\author[orcid=0009-0002-8461-6111,sname='Pan']{Yanbo Pan (潘彦博)}
\altaffiliation{UVa Provost Fellow}
\affiliation{Department of Astronomy, University of Virginia, 530 McCormick Road, Charlottesville, VA 22904, USA}
\affiliation{Department of Astronomy, University of Michigan, 1085 S. University Ave., Ann Arbor, MI 48109, USA}
%\email[show]{panpi@umich.edu}
\email[show]{war8sk@virginia.edu}

\author[0000-0003-4816-3469]{Ryan J. MacDonald}
\altaffiliation{NHFP Sagan Fellow}
\affiliation{School of Physics and Astronomy, University of St Andrews, North Haugh, St Andrews, KY16 9SS, UK}
\affiliation{Department of Astronomy, University of Michigan, 1085 S. University Ave., Ann Arbor, MI 48109, USA}
\email[]{Ryan.MacDonald@st-andrews.ac.uk}

\author[0000-0001-9513-1449]{N\'estor Espinoza}
\affiliation{Space Telescope Science Institute, 3700 San Martin Drive, Baltimore, MD 21218, USA}
\affiliation{William H. Miller III Department of Physics and Astronomy, Johns Hopkins University, Baltimore, MD 21218, USA}
\email[]{nespinoza@stsci.edu}

\author[0000-0001-7216-4846]{Bibiana Prinoth}
\altaffiliation{ESO fellow}
\affiliation{European Southern Observatory, Karl-Schwarzschild-Strasse 2, 85748 Garching, Germany}
\affiliation{Lund Observatory, Division of Astrophysics, Department of Physics, Lund University, Box 43, 221 00 Lund, Sweden}
\email[]{bibiana.prinoth@eso.org}

\author[0000-0003-4408-0463]{Zafar Rustamkulov}
\affiliation{Department of Earth and Planetary Science, Johns Hopkins University, 3400 N. Charles Street, Baltimore, MD 21218, USA}
\email[]{zafar@jhu.edu}

\author[0000-0003-0156-4564]{Luis Welbanks}
\affiliation{School of Earth and Space Exploration, Arizona State University, 781 Terrace Mall, Tempe, AZ 85287, USA}
\email{luis.welbanks@asu.edu}

\author[0000-0003-4269-3311]{Daniel Kitzmann}
\affiliation{Space Research and Planetary Sciences, Physics Institute, University of Bern, Gesellschaftsstrasse 6, 3012 Bern, Switzerland} 
\affiliation{Center for Space and Habitability, University of Bern, Gesellschaftsstrasse 6, 3012 Bern, Switzerland}
\email[]{daniel.kitzmann@unibe.ch}  

\author[orcid=0000-0001-7866-8738]{Nicolas Crouzet}
\affiliation{Kapteyn Astronomical Institute, University of Groningen, P.O. Box 800, 9700 AV Groningen, The Netherlands}
\affiliation{Leiden Observatory, Leiden University, P.O. Box 9513, 2300 RA Leiden, The Netherlands}
\email[]{crouzet@astro.rug.nl}

\author[0000-0001-6960-0256]{Anna Walde}
\affiliation{University Observatory Munich, Ludwig Maximilian University, Scheinerstrasse 1, Munich D-81679, Germany} 
\affiliation{Center for Space and Habitability, University of Bern, Gesellschaftsstrasse 6, 3012 Bern, Switzerland}
\email[]{anna.lueber@unibe.ch}  

\author[orcid=0000-0002-4250-0957]{Diana Powell}
\affiliation{Department of Astronomy \& Astrophysics, University of Chicago, 5640 South Ellis Avenue, Chicago, IL 60637, USA}
\email[]{diana.powell@uchicago.edu}

\author[0000-0003-0973-8426]{Eva-Maria Ahrer}
\affiliation{Max Planck Institute for Astronomy, K\"onigstuhl 17, D-69117 Heidelberg, Germany}
\email[]{ahrer@mpia.de}

\author[0000-0002-6523-9536]{Adam J. Burgasser}
\affiliation{UCSD Department of Astronomy \& Astrophysics, 9500 Gilman Drive, La Jolla, CA 92093, USA}
\email[]{aburgasser@ucsd.edu}

\author[0000-0002-4997-0847]{Duncan A. Christie}
\affiliation{Max Planck Institute for Astronomy, K\"onigstuhl 17, D-69117 Heidelberg, Germany}
\email[]{christie@mpia.de}

\author[0000-0002-8658-3811]{Wolf Cukier}
\affiliation{Department of Astronomy \& Astrophysics, University of Chicago, 5640 South Ellis Avenue, Chicago, IL 60637, USA}
\email[]{wcukier@uchicago.edu}

\author[0000-0001-5097-9251]{Carlos Gasc\'on}
\affiliation{Center for Astrophysics | Harvard \& Smithsonian, 60 Garden Street, Cambridge MA 02138, USA}
\affiliation{Institut d'Estudis Espacials de Catalunya (IEEC), 08860 Castelldefels, Barcelona, Spain}
\email[]{carlos.gascon@cfa.harvard.edu}

\author[0000-0002-0931-735X]{M\r{a}ns Holmberg}
\affiliation{Space Telescope Science Institute, 3700 San Martin Drive, Baltimore, MD 21218, USA}
\email[]{mholmberg@stsci.edu}  

\author[0000-0002-2984-3250]{Thomas Kennedy}
\affiliation{Department of Astronomy, University of Michigan, 1085 S. University Ave., Ann Arbor, MI 48109, USA}
\email[]{thomak@umich.edu}

\author[0000-0003-3204-8183]{Mercedes L\'opez-Morales}
\affiliation{Space Telescope Science Institute, 3700 San Martin Drive, Baltimore, MD 21218, USA}
\email[]{mlopez-morales@cfa.harvard.edu} 

\author[0000-0001-6707-4563]{N. J. Mayne}
\affiliation{Department of Physics and Astronomy, Faculty of Environment Science and Economy, University of Exeter, EX4 4QL, UK.}
\email[]{n.j.mayne@exeter.ac.uk}
 
\author[0000-0002-8956-2047]{Dominic Samra}
\affiliation{University of Chicago, William Eckhardt Research Center, 5640 South Ellis Avenue, Chicago, IL 60637, USA }
\email{dominicsamra.uk@gmail.com}

\author[0000-0002-2454-768X]{Arjun B. Savel}
\affiliation{Astronomy Department, University of Maryland, College Park, 4296 Stadium Drive, College Park, MD 207842, USA}
\email[]{asavel@umd.edu}

\author[0000-0001-8342-1895]{Maria E. Steinrueck}
\altaffiliation{51 Pegasi b fellow}
\affiliation{Department of Astronomy \& Astrophysics, University of Chicago, 5640 South Ellis Avenue, Chicago, IL 60637, USA}
\email[]{msteinrueck@uchicago.edu}

\author[0000-0003-1656-011X]{Christopher P. Wirth}
\affiliation{Department of Astronomy \& Astrophysics, University of Chicago, 5640 South Ellis Avenue, Chicago, IL 60637, USA}
\email[]{cwirth@uchicago.edu}

\collaboration{all}{JWST GO 3969 Team}

\begin{abstract}

Exoplanets orbiting active stars offer distinct challenges for atmospheric characterization with the JWST. Among such worlds, the hot Saturn WASP-52b ($T_{\rm{eq}} \sim$ 1300\,K) orbits an active K-dwarf that complicates transmission spectroscopy via stellar contamination. Here, we present the first JWST NIRSpec G395H (2.8-5.2\,$\micron$) limb-averaged transmission spectrum of WASP-52b obtained through the JWST Morning/Evening Program (GO-3969), which aims to measure limb asymmetries across a sample of hot giant exoplanets. Our spectrum reveals the first detection of \ce{CO2} in WASP-52b's atmosphere ($\ln \mathcal{B} = 54$ / $> 10\,\sigma$), with a retrieved abundance of log $\rm C O_2 = -5.09^{+1.09}_{-1.14}$, and confirmation of \ce{H2O} ($\ln \mathcal{B} = 5.9$ / $\sim$ 3--4$\,\sigma$), with an abundance of log $\rm H_2 O = -3.41^{+1.04}_{-1.04}$. We find a steep near-infrared slope that cannot be explained solely by \ce{H2O} absorption, necessitating unocculted starspots ($\ln \mathcal{B} = 5.7$ / $\sim 3-4\,\sigma$), though with properties that vary across our data reductions. The shape and amplitude of the molecular bands additionally suggest high-altitude ($<$ 10\,mbar) inhomogeneous clouds covering $\approx 40 \pm 20$\% of the terminator ($\ln \mathcal{B} = 1.9$ / $\sim 2.5\,\sigma$), which provides evidence for the  morning-evening terminator differences that are consensus predictions of general circulation models. WASP-52b's atmospheric metallicity can be sub-solar, solar, or super-solar, due to order-of-magnitude uncertainties in the molecular abundances caused by degeneracies between compositions, clouds, and starspots. However, chemical equilibrium retrievals provide a similar statistical fit while favoring a tighter constraint with a super-solar metallicity (M/H $= 14.8^{+12.1}_{-5.9} \times$ solar). These results highlight that chemical detections and aerosol properties may still be recovered for planets orbiting active stars via transmission spectroscopy.

\end{abstract}

\keywords{exoplanet astronomy: extrasolar gaseous planets, exoplanet structure: exoplanet atmospheres, stellar physics: starspots, astronomical techniques: transmission spectroscopy}

\section{Introduction}
\label{sec:intro}
%%%%% Hot Jupiter with JWST, highlights big past discoveries %%%%%
The field of exoplanet atmospheres is undergoing a renaissance, largely due to the increasing sample of exoplanet transmission spectra from JWST \citep[see, e.g.,][for a review]{jwst-exoreview}. Transmission spectra \citep{Seager2000, Brown2001} probe the terminator region of transiting exoplanets, and thus are sensitive to the atmospheric properties around the day-night boundary \citep[e.g.][]{kempton2017,powell2019,espinoza2021, wardenier2022, MacDonald2022}. However, one important caveat of transmission spectroscopy is that active stars---with non-uniform surface features such as cool spots---can also imprint spectral signatures of a non-planetary origin \citep[see][for a review]{rackham2023, Moran2023}. 

Giant exoplanets are predicted to have significant differences in their temperature profiles, chemical abundances, and aerosol properties between their morning and evening terminators, according to 3-dimensional general circulation models (GCMs) \citep[e.g.][]{Showman2002,Dobbs-Dixon2008,Mayne2017,helling2019wasp18b,helling2019,helling2023}. These predictions have manifested in early JWST spectra of hot Jupiters. The hot Jupiter WASP-39b ($T_{\rm{eq}} \approx 1200$\,K; \citealt{Faedi2011}) provided evidence of inhomogeneous (`patchy') clouds in one of the first JWST transmission spectra \citep{feinstein2023,constantinou2023}, with subsequent analyses showing distinct morning and evening terminator spectra \citep{Espinoza2024} which imply a hotter evening terminator with different aerosol properties on each limb \citep{Steinrueck2025,Chen2025}. The comparatively cooler planet WASP-107b ($T_{\rm{eq}} \approx 770$\,K; \citealt{Anderson2017}) also shows asymmetric terminator spectra \citep{Murphy2024} driven by a combination of temperature, aerosol, and CO$_2$ and SO$_2$ abundance differences between the morning and evening terminators \citep{fu2025, mukherjee2025, ahrer2025}. 

A detailed analysis of the origin and prevalence of morning-evening terminator differences requires a sample of JWST transmission spectra for planets predicted to exhibit asymmetric terminators. To this end, we are conducting the JWST Morning/Evening Program (MEP; JWST GO-3969; co-PIs: Espinoza \& Powell) to provide the first systematic survey aimed at constraining inhomogeneous limbs for giant exoplanets. Our survey spans planets from 1000--2000\,K, including both archival planets and new JWST observations of three planets: WASP-52b ($T_{\rm{eq}} \approx 1300$\,K; \citealt{Hebrard2013}), Kepler-12b ($T_{\rm{eq}} \approx 1500$\,K; \citealt{Fortney2011}), and HAT-P-65b ($T_{\rm{eq}} \approx 1900$\,K; \citealt{Hartman2016}). We constructed our sample to constrain terminator temperature differences and C/O ratios to roughly the same degree across these exoplanets, enabling direct tests of predictions from cloud formation models and GCMs.

In this paper, we present the first observation from the JWST MEP: the NIRSpec G395H limb-averaged transmission spectrum of WASP-52b. WASP-52b is a hot Jupiter ($\rm R_p = 1.27R_J$, $\rm M_p = 0.46M_J$, $P=1.7\text{ day}$; \citealt{Hebrard2013}) orbiting an active K-dwarf star ($\rm T_{eff}=5008K$). With its high temperature ($\rm T_{eq} = 1315$\,K), large scale height ($\rm H\simeq$700\,km), and low surface gravity (log g$\rm _p = 2.81 cm s^{-2}$), WASP-52b serves as an ideal target with significant transit depth ($\rm \delta = 2.71\%$) for the transmission spectroscopy technique.

WASP-52b's atmosphere has been studied for many years using transmission spectroscopy from ground- and space-based telescopes. Early low-resolution optical spectra of WASP-52b showed a relatively muted transmission spectrum attributed to high-altitude clouds \citep{kirk2016,Louden2017} with Na absorption \citep{chen2017}. The cloudiness of WASP-52b was confirmed by Hubble (HST)/STIS optical observations and Spitzer observations \citep{alam2018}, with subsequent near-infrared HST/WFC3 observations showing muted water vapor features  \citep{Bruno2018HST}. Subsequent high-resolution ground-based transmission spectroscopy with VLT/ESPRESSO confirmed the presence of Na and additionally detected K in WASP-52b's atmosphere \citep{Chen2020}. Throughout these early analyses, a consistent theme was attempts to model and/or correct for the influence of occulted and unocculted stellar active regions of WASP-52b's transmission spectrum \citep{kirk2016,alam2018,Bruno2018HST,bruno2020}.

Recently, \cite{fournier_tondreau2025} presented the first JWST transmission spectrum of WASP-52b using the NIRISS/SOSS instrument (0.6 - 2.8 $\micron$). These observations revealed significant spectral contamination from both occulted and unocculted stellar active regions, atmospheric H$_2$O ($\log$\,H$_2$O $\approx -4 \pm 1$), escaping He, and hints of K, alongside the presence of scattering hazes. This study found that a model including unocculted starspots and faculae is preferred at 3.6$\sigma$ compared to an atmosphere-only model, with unocculted starspots/faculae covering $\approx30\pm10$\%/$\approx20\pm10$\% of the stellar surface, respectively. However, the wavelength coverage of these NIRISS/SOSS observations offered little sensitivity to other near-infrared absorbers, such as CO$_2$, CH$_4$, HCN, SO$_2$, and H$_2$S. The JWST NIRSpec/G395H spectrum therefore complements the NIRISS/SOSS spectrum by enabling constraints on additional near-infrared chemical trace species. The simultaneous presence of atmospheric absorption, significant stellar contamination, and scattering aerosols renders WASP-52b an ideal laboratory to assess the influence of the transit light source effect \citep{Rackham2018,Rackham2019} for a giant planet orbiting a K-dwarf.

% Many studies have probed WASP-52b's atmosphere using ground-based high-resolution spectroscopy. \cite{nail2025} recently showed WASP-52b's atmospheric outflow represents a transitional stage between a tail from hot outflows and stream-like morphology from cold outflows, using high-resolution spectra with CRIRES$^{+}$ and CARMENES. \cite{chen2017} detected the sodium doublet narrow absorption feature from 522nm to 903nm, using low-Intermediate-Resolution Integrated Spectroscopy (OSIRIS) at the 10.4 m Gran Telescopio Canarias (GTC). 

This paper aims to examine WASP-52b's NIRSpec/G395H spectrum to characterize its chemical inventory, aerosol properties, and stellar contamination of its host star. Our WASP-52b JWST NIRSpec/G395H (2.8-5.2 $\mu$m) transmission spectrum analysis is structured as follows. In Section~\ref{sec:data}, we describe the JWST transit observation, data reduction by \texttt{transitspectroscopy}, \texttt{FIREFly}, and the \texttt{Default JWST} pipeline, in addition to the light-curve fitting process. In Section~\ref{sec:retrievals}, we present our POSEIDON and Aurora retrieval analysis followed by atmospheric/stellar interpretation. Finally, we summarize and discuss our results in Section \ref{sec:discu}.

\section{Observations \& Data Reduction} \label{sec:data}

% Program, PI, time, instrument
% See GO proposal % https://www.stsci.edu/jwst/science-execution/program-information?id=3969

A transit of WASP-52b was observed using the G395H mode of the NIRSpec instrument as part of the JWST General Observers (GO) program PID: 3969 (PI: N\'estor  Espinoza \& Diana Powell). The time series observation (TSO) started on Oct 30$^\mathrm{th}$, 2023 23:26 UT and spanned 5.5 hours, ending on Oct 31$^\mathrm{st}$, 2023 04:59. This covered the 2-hour transit of WASP-52~b, with a pre-transit baseline of 2.5 hours and a post-transit baseline of 1 hour.

\begin{figure*}
    \centering\includegraphics[width=1.0\linewidth]{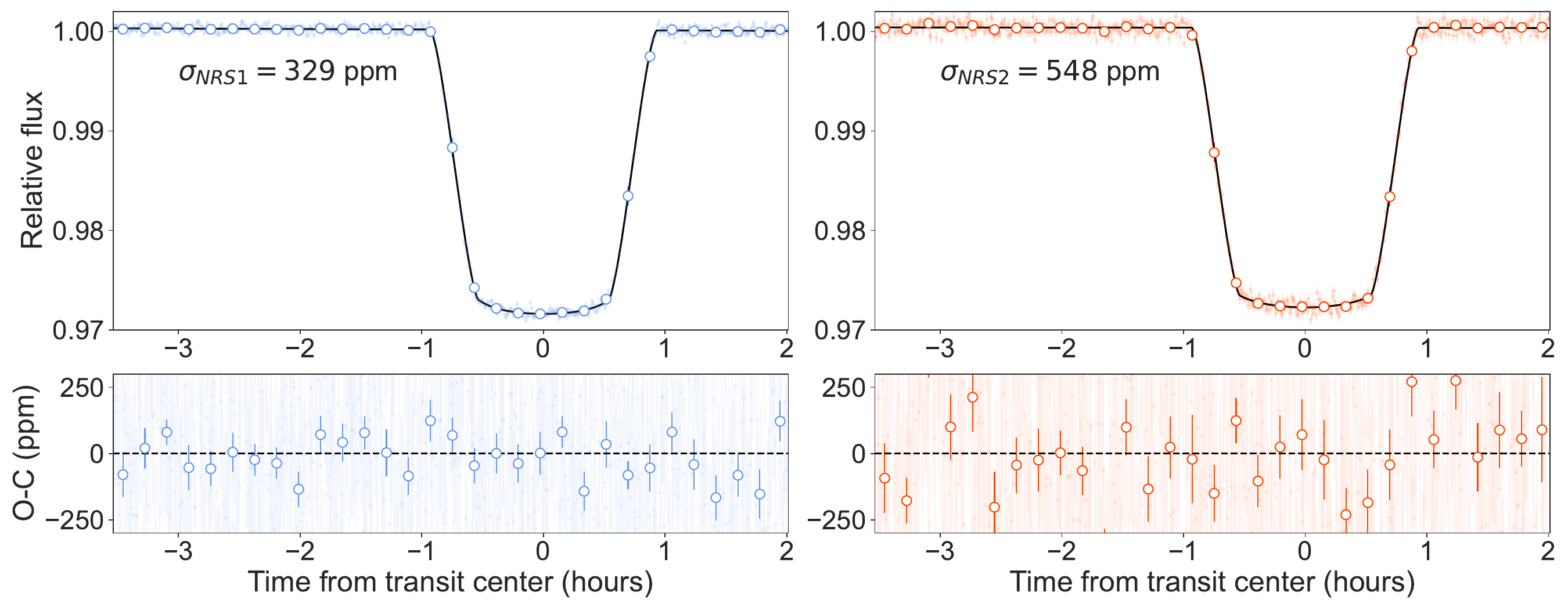}
    \caption{JWST NIRSpec/G395H white-light transits of WASP-52b by \textsc{transitspectroscopy}. The light curves showcase the precision obtained by the light curves captured in the NRS1 and NRS2 detectors. We see no evidence in the residuals of spot crossing events up to about 100 ppm levels.}
    \label{fig:white-light}
\end{figure*}

We reduce the data using three independent pipelines: the \texttt{transitspectroscopy} pipeline as described in \cite{Espinoza:2025}, the \texttt{FIREFLy} pipeline described in \cite{Rustamkulov2023} and a data reduction using the \textit{JWST} notebooks published by STScI on JWebbinar 29\footnote{\url{https://www.stsci.edu/jwst/science-execution/jwebbinars}} which we refer to as the \texttt{Default JWST} data reduction below. A white-light light curve for both NRS1 and NRS2 detectors using the \texttt{transitspectroscopy} pipeline is presented in Figure \ref{fig:white-light}, which showcases the exquisite data quality of our observations, on which we observe no spot crossing events.

\begin{deluxetable*}{lccc}
\tablecaption{Priors and Posterior Ephemerides and Orbital Parameters for WASP-52~b using broad-band ``white light" curves. Prior values from \citep{kokori:2023} \label{tab:params}}
\tablehead{
\colhead{Parameter} & \colhead{Prior} & \colhead{Posterior Value} & \colhead{Unit}
}
\startdata
$P$                        & Fixed: 1.74978117                    & ---                                            & days \\
$T_0$                      & $\mathcal{N}(2460248.625, 0.2)$       & $2460248.6240_{-0.000018}^{+0.000018}$         & days \\
$a/R_\star$                & $\mathcal{N}(7.38, 0.5)$              & $7.2190_{-0.0199}^{+0.0197}$                   & --- \\
$b$                        & $\mathcal{N}_\mathrm{trunc}(0.6,0.1;0.0,1.0)$ & $0.6007_{-0.0039}^{+0.0038}$          & --- \\
$e$                        & Fixed: 0.0                            & ---                                            & --- \\
$\omega$                   & Fixed: 90.0                           & ---                                            & deg \\
$R_p/R_s\,$(NRS1)           & $\mathcal{U}(0.0, 0.2)$               & $0.1667_{-0.0003}^{+0.0002}$                   & --- \\
$R_p/R_s\,$(NRS2)           & $\mathcal{U}(0.0, 0.2)$               & $0.1659_{-0.0003}^{+0.0003}$                   & --- \\
\enddata
\tablecomments{Priors are shown as distributions: $\mathcal{N}(\mu,\sigma)$ for normal, $\mathcal{U}(a,b)$ for uniform, and $\mathcal{N}_\mathrm{trunc}(\mu,\sigma;a,b)$ for truncated normal with mean $\mu$, standard deviation $\sigma$, and bounds $[a,b]$. Fixed parameters were not varied during the fit. Posteriors are reported as median with 1$\sigma$ uncertainties.}
\end{deluxetable*}

\subsection{\texttt{transitspecroscopy} data reduction}

For the \texttt{transitspectroscopy} data reduction, we follow the same configuration outlined in \cite{Espinoza:2025}, except that we use the standard saturation reference file from the JWST Calibration Pipeline \citep[][version 1.10.2]{jwst-calibration-pipeline}. We do perform an additional 1/f correction at the rates per integration level using the group-scaling technique described in \cite{albert:2023}. In short, this scales the median out-of-transit rates per integrations to each individual integration, subtracts it, and then the residual image is used to estimate the 1/f signal at each column. To this end, the median of the 10 pixels closest to the central trace on this residual image are used as such estimate, and thus subtracted from the original rates per integration.

Using this reduction, we form a band-integrated, ``white light" curve for both NRS1 and NRS2 by extracting the spectra with a 3-pixel aperture radius. We jointly fit these light curves with \texttt{juliet} \citep{juliet} to obtain updated ephemeris and orbital parameters for WASP-52~b. These are presented in Table \ref{tab:params}. For this fit, we simply fit the light curves with a time-dependent slope as our systematics model \citep[known to be different for NRS1 and NRS2]{Espinoza:2023}, free quadratic limb-darkening using the parametrization of \cite{Kipping:2013}, and an extra jitter term --- all different between NRS1 and NRS2. We also fit a different $R_P$/$R_{\star}$ for each detector. As can be observed in the models and residuals in Figure \ref{fig:white-light}, this model is a very good fit to the data, with an overall RMS per point of 329 and 548 ppm for NRS1 and NRS2, respectively. The resulting orbital parameters are then used in the wavelength-dependent fits, which are performed at the resolution level of the instrument. For this reduction, the wavelength-dependent fits use the same setup, but fixing the ephemerids and orbital parameters of the system to those of Table \ref{tab:params}.

\subsection{\texttt{FIREFLy} data reduction}

The \texttt{FIREFLy} data reduction uses version 3.7 to perform the default Stage 1 from the JWST Calibration Pipeline of \cite{jwst-calibration-pipeline}, with 1/f noise removed at the group-level stage, and skipping the jump step. To trace the spectra, a gaussian was fitted to each columns, and the centers of those gaussians were all fitted with a low-order polynomial to remove outleirs. To extract the spectrum, and aperture radius of 3 pixels was used. For light curve fitting, the wavelength-dependent fits were performed similarly to the \texttt{transitspectroscopy} reduction, with a linear trend fitted to the data, and limb darkening also allowed to freely fit using the same priors and setup as with the \texttt{transitspectroscopy} pipeline reduction described above.

\subsection{\texttt{Default JWST} data reduction}

This pipeline made use of the notebooks in JWebbinar 29 with small modifications. In particular, version 1.14.0 of the JWST Calibration Pipeline is used. Spectral extraction using these notebooks use an aperture radius of 2 pixels. Then, light curves as a function of wavelength are fit similarly to the previous reductions.

\subsection{Transmission Spectrum} \label{subsec:spectra}

\begin{figure*}
    \centering\includegraphics[width=1.0\linewidth]{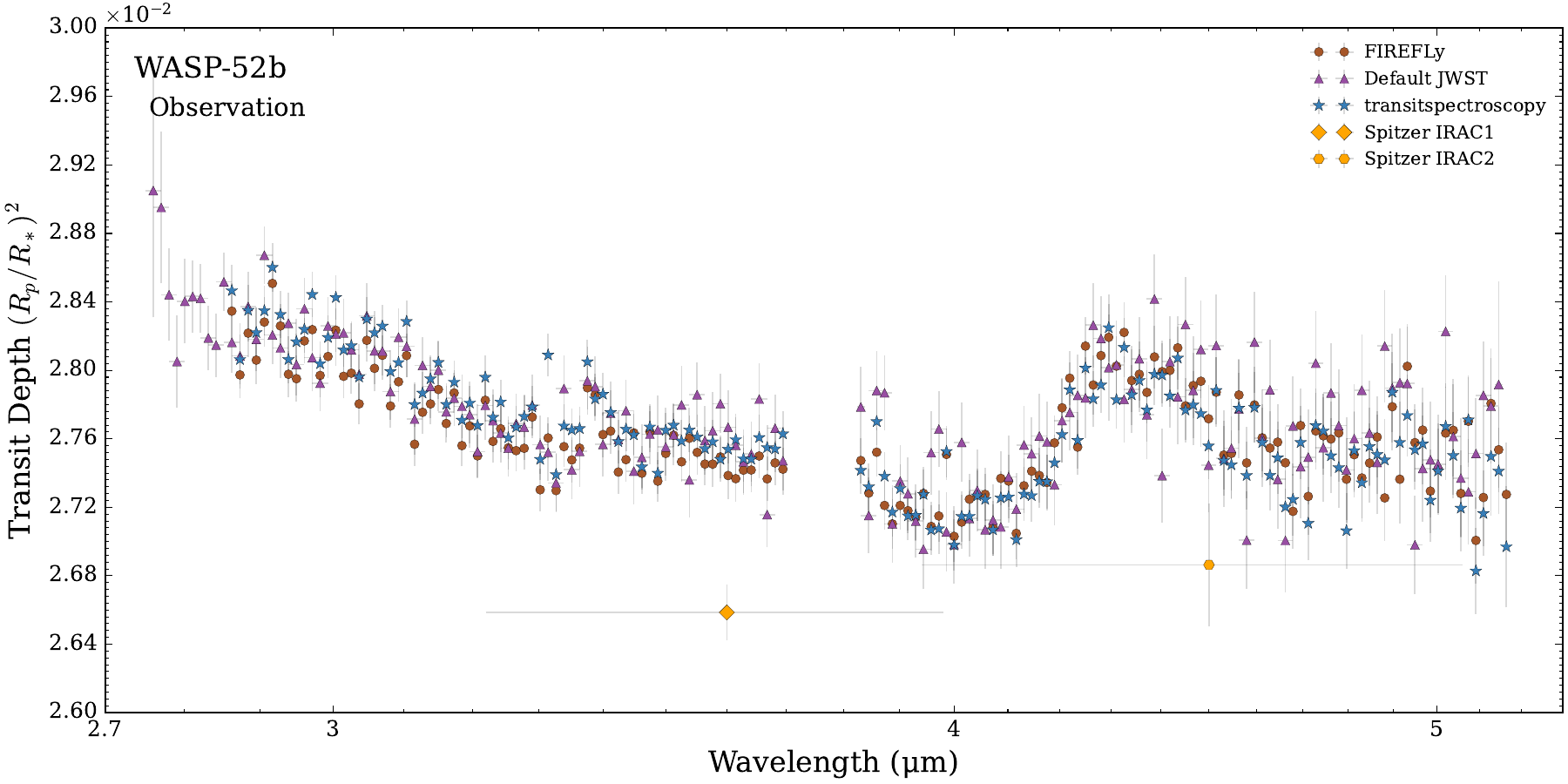}
    \caption{JWST NIRSpec/G395H transmission spectrum of WASP-52b. Three independent data reductions are compared: \texttt{FIREFly} (brown circles), the JWST default pipeline (purple triangles), and \texttt{transitspectroscopy} (blue stars). The Spitzer/IRAC data from \cite{alam2018} are overplotted for comparison for the 3.6 $\mu$m and 4.5 $\mu$m filters (orange diamonds).}
    \label{fig:obs}
\end{figure*}

We compare our NIRSpec/G395H transmission spectra of WASP-52b from \texttt{transitspectroscopy}, \texttt{FIREFly}, and the \texttt{Default JWST} pipelines in Figure~\ref{fig:obs}. The spectrum is characterized by two prominent features: (i) a slope with increasing transit depth towards short wavelengths (attributable to H$_2$O and/or unocculted starspots); and (ii) the CO$_2$ absorption band centered on 4.3\,\micron. Our data are significantly offset from the previous Spitzer/IRAC observations \citep{alam2018}. To investigate this JWST--Spitzer offset, we repeated the \texttt{transitspectroscopy} JWST reduction using the Spitzer-derived orbital parameters from \citet{alam2018}. This creates a uniform 124\,ppm offset, which is insufficient to explain the $\approx 950$\,ppm offset at 3.6\,$\mu$m and $\approx 700$\,ppm offset at 4.5\,$\mu$m between the Spitzer and JWST/NIRSpec data (see Section \ref{appendix:Spitzer} for more details). This suggests the offset arises from a changing baseline from stellar variability between the epochs of the Spitzer and JWST observations. We now turn to our quantitative analysis of WASP-52b's transmission spectrum.

\section{Atmospheric Retrieval Analysis} \label{sec:retrievals}

We analyze the NIRSpec G395H transmission spectrum of WASP-52b using two Bayesian atmospheric retrieval frameworks: POSEIDON \citep{macDonald2017,macDonald2023} and Aurora \citep{Welbanks2021}. These two codes both consider the spectroscopic signatures of a planetary atmosphere as well as an inhomogeneous stellar photosphere. Below we describe each atmospheric retrieval approach and their resulting interpretation for the nature of WASP-52b.

%Modified to describe Poseidon
\subsection{Atmospheric Inference using \textsc{Poseidon}} \label{subsec:poseidon_methods}

We perform retrievals on WASP-52\,b's transmission spectrum using the open-source atmospheric retrieval codes \textsc{Poseidon} \citep{macDonald2017,macDonald2023}. All retrievals are conducted with 1,000 MultiNest live points to explore the posterior distributions of the multidimensional parameter space smoothly \citep{feroz2009}.

For Bayesian model comparisons, we provide both Bayes factors ($\mathcal{B}_{01}$) between two models (e.g. for models with and without CO$_2$) and an upper limit on the equivalent `detection significance' using the relation from \citet{Benneke2013}. We note that ongoing discussions in the exoplanet literature generally recommend quoting Bayes factors for the more conservative statistical preference between two models \citep{Kipping2025,Thorngren2026}.

Our \textsc{Poseidon} atmospheric models use a six-parameter pressure-temperature (P-T) profile following the \citet{madhu2009} prescription with a reference temperature parameter at 10\,mbar. We assume a hydrogen-helium dominated composition with a fixed \ce{He}/\ce{H2} ratio of 0.17. The modeled atmosphere extends from $10^{-7}$ to $10^2$\,bar and is discretized into 100 layers uniformly spaced in log pressure. Hydrostatic equilibrium is anchored at a reference pressure of 10 bar as the boundary condition, where the retrieved reference radius is located. We also test isothermal P-T profile and find retrievals with isothermal profiles give consistent results comparing to those following \citet{madhu2009}. Regardless, we marginalize over P-T profiles to ensure unbiased results. Aerosols are included using the four-parameter inhomogeneous cloud and haze treatment of \citet{macDonald2017}.

We include several trace chemical species that are expected to exist at WASP-52\,b's equilibrium temperature \citep{Madhusudhan2016,Woitke2018,Mukherjee2024} and produce prominent absorption features within the NIRSpec/G395H wavelength range: \ce{H2O}, \ce{CH4}, \ce{CO}, \ce{CO2}, \ce{H2S}, \ce{HCN}, \ce{SO2}, and \ce{K}. Our opacities are sourced from the \textsc{POSEIDON} v1.2 release \citep{Mullens2024}, incorporating state-of-the-art line lists and pressure broadening parameters from: \ce{H2O} \citep{Polyansky2018}, \ce{CH4} \citep{Yurchenko2024}, \ce{CO} \citep{li2015}, \ce{CO2} \citep{Yurchenko2020}, \ce{H2S} \citep{azzam2016}, \ce{HCN} \citep{Barber2014}, \ce{SO2} \citep{underwood2016}, and \ce{K} \citep{Ryabchikova2015}. Cross-sections are computed using the open-source \texttt{Cthulhu} Python package \citep{Agrawal2024}. Additional opacity sources include Rayleigh scattering by \ce{H2} \citep{Hohm1994} and collision-induced absorption from \ce{H2}-\ce{H2} and \ce{H2}-\ce{He} pairs \citep{Karman2019}. 

In light of unocculted stellar features identified in WASP-52\,b's previous transmission spectra from \textit{HST} \citep{bruno2020} and \textit{JWST} \citep{fournier_tondreau2025}, we also include stellar contamination parameters in our retrievals. We use two stellar contamination setups, following \citet{fournier_tondreau2024, fournier_tondreau2025}: (i) an atmosphere-only (i.e. stellar contamination-free) model and (ii) a one-heterogeneity (starspots) + atmosphere model. The second one adds three parameters: the stellar photosphere temperature, the heterogeneity temperature (less than the photosphere for spots, greater than the photosphere for faculae), and the heterogeneity coverage fraction. We tested retrieving different surface gravities for the heterogeneities compared to the photosphere as in \citealt{fournier_tondreau2024}, but we found this to be unnecessary. We also included the faculae contribution in our test retrievals, but found it is not statistically significant, since faculae have minimal influence on the long-wavelength NIRSpec/G395H data. We compute the stellar contamination contribution factor by interpolating PHOENIX models \citep{husser2013} using the PyMSG package \citep{Townsend2023}. We apply both models to the \texttt{transitspectroscopy}, \texttt{FIREFly}, and \texttt{Default JWST} reduced transmission spectra shown in Figure \ref{fig:obs}.

We calculate model spectra at a resolution of $R = \lambda/d\lambda = 20,000$ from 0.58 to 6.00\,$\mu$m using the configuration described above. The model spectra are calculated via opacity sampling onto this intermediate resolution wavelength grid from the high-resolution opacities ($\Delta\nu$ = 0.01 cm$^{-1}$, equivalent to $R = \lambda / \Delta\lambda = 10^6$ at 1\,$\mu$m). We additionally include a relative offset parameter, $\delta_{\rm rel}$, to account for the vertical offset between NIRSpec/G395H NRS 1 and NRS 2 spectra. Considering the atmospheric properties, stellar contamination, and NRS 2 vs. NRS 1 offset, the POSEIDON retrieval models have the following number of free parameters: 20 for the atmosphere-only model, and 23 for the one-heterogeneity + atmosphere model. We additionally tested including an error inflation factor in our retrievals \citep{Line2015}, but found no notable effect on our posterior widths so this parameter was excluded from our final retrieval configuration. We did not impose strong assumptions in our prior configurations. Most parameter priors are uniform except for the stellar photosphere temperature, for which we adopt a previous constraint from \cite{Hebrard2013}. We summarize the priors for each parameter in Table~\ref{tab:retrieval_priors}.

\begin{table*}[htbp]
\centering
\caption{Retrieval Priors}
\label{tab:retrieval_priors}
% \begin{tabular}{l>{\raggedright\arraybackslash}p{4.5cm}*{4}{>{\centering\arraybackslash}p{1.8cm}} @{}}
\begin{tabular}{lp{4.5cm}cccc}
\toprule

 &  & \multicolumn{2}{c}{POSEIDON} & \multicolumn{2}{c}{Aurora} \\
\cmidrule(lr){3-4} \cmidrule(lr){5-6}
Parameter & Description  & Atmosphere & Atmosphere & Atmosphere & Atmosphere \\
 & & Only & +Spots & Only & +Spots \\
\midrule
\textbf{P-T Profile}  & & & & & \\
$T_{\textrm{ref}}$ (K) & Reference temperature & $\mathcal{U}(300,1600)$ & $\mathcal{U}(300,1600)$ & $\mathcal{U}(800,1600)$ & $\mathcal{U}(800,1600)$ \\
$\alpha_1$ & P-T profile curvature & $\mathcal{U}(0.02,2.0)$ & $\mathcal{U}(0.02,2.0)$ & $\mathcal{U}(0.02,2.0)$ & $\mathcal{U}(0.02,2.0)$ \\
$\alpha_2$ & P-T profile curvature & $\mathcal{U}(0.02,2.0)$ & $\mathcal{U}(0.02,2.0)$ & $\mathcal{U}(0.02,2.0)$ & $\mathcal{U}(0.02,2.0)$ \\
$\log(P_1/\textrm{bar})$ & P-T profile region boundary & $\mathcal{U}(-7,2)$ & $\mathcal{U}(-7,2)$ & $\mathcal{U}(-9,2)$ & $\mathcal{U}(-9,2)$ \\
$\log(P_2/\textrm{bar})$ & P-T profile region boundary & $\mathcal{U}(-7,2)$ & $\mathcal{U}(-7,2)$ & $\mathcal{U}(-9,2)$ & $\mathcal{U}(-9,2)$ \\
$\log(P_3/\textrm{bar})$ & P-T profile region boundary & $\mathcal{U}(-2,2)$ & $\mathcal{U}(-2,2)$ & $\mathcal{U}(-2,2)$ & $\mathcal{U}(-2,2)$ \\

\midrule
\textbf{Chemical Composition} & & & & & \\
$\log(X_i)$ & Chemical mixing ratios & $\mathcal{U}(-12,-1)$ & $\mathcal{U}(-12,-1)$ & $\mathcal{U}(-12,-1)$ & $\mathcal{U}(-12,-1)$ \\

\midrule
\textbf{Aerosols} & & & & & \\
$\log(a)$ & Rayleigh enhancement factor & $\mathcal{U}(-4,8)$ & $\mathcal{U}(-4,8)$ & $\mathcal{U}(-4,8)$ & $\mathcal{U}(-4,8)$ \\
$\gamma$ & Scattering slope & $\mathcal{U}(-20,2)$ & $\mathcal{U}(-20,2)$ & $\mathcal{U}(-20,2)$ & $\mathcal{U}(-20,2)$ \\
$\log(P_{\textrm{cloud}}/\textrm{bar})$ & Cloud top pressure & $\mathcal{U}(-6,2)$ & $\mathcal{U}(-6,2)$ & $\mathcal{U}(-9,-2)$ & $\mathcal{U}(-9,-2)$ \\
$\phi$ & Cloud/Haze coverage fraction & $\mathcal{U}(0,1)$ & $\mathcal{U}(0,1)$ & $\mathcal{U}(0,1)$ & $\mathcal{U}(0,1)$ \\

\midrule
\textbf{Stellar} & & & & & \\
$T_{\textrm{phot}}$ (K) & Photosphere temperature & — & $\mathcal{N}(5000,100)$ & — & $\mathcal{N}(5000,100)$ \\
$f_{\textrm{het}}$ (K) & Heterogeneity coverage fraction & — & $\mathcal{U}(0.0,0.5)$ & — & $\mathcal{U}(0.0,0.5)$ \\
$T_{\textrm{het}}$ (K) & Heterogeneity temperature & — & $\mathcal{U}(3500,6000)$ & — & $\mathcal{U}(3500,6000)$ \\

\midrule
\textbf{Other Parameters} & & & & & \\
$R_{\textrm{P,ref}}$ ($R_{\textrm{Jup}}$) & Reference planet radius & $\mathcal{U}(0.89,1.46)$ & $\mathcal{U}(0.89,1.46)$ & $\mathcal{U}(0.89,1.46)$ & $\mathcal{U}(0.89,1.46)$ \\
$\delta_{\textrm{rel}}$ (ppm) & NIRSpec NRS2 offset & $\mathcal{U}(-500,500)$ & $\mathcal{U}(-500,500)$ & $\mathcal{U}(-500,500)$ & $\mathcal{U}(-500,500)$ \\
\bottomrule
\end{tabular}
\end{table*}

\subsection{Atmospheric Inference using Aurora} \label{subsec:aurora_methods}

We complement the analysis from \textsc{Poseidon} above with a second independent analysis using Aurora \cite{Welbanks2021}. Briefly, Aurora is a tool for the analysis of transmission \citep[e.g.,][]{Welbanks2024} and emission \citep[e.g.,][]{Bell2023} spectra of exoplanet atmospheres. An atmospheric model with a parametrized chemical composition, aerosol properties, and vertical temperature structure is coupled with a parameter estimation method; in this case MultiNest \citep{feroz2009} through PyMultiNest \citep{Buchner2014}. The analysis with Aurora considers the effects of an inhomogeneous stellar photosphere during the primary transit giving rise to the so-called Transit Light Source effect \citep[e.g.,][]{Rackham2018} using the prescription introduced in \cite{Pinhas2018}.

The model atmosphere closely follows the implementation in \textsc{Poseidon} to enable a close comparison between methodologies and their inferred atmospheric properties. The model follows a plane-parallel atmosphere under hydrostatic equilibrium divided in 100 pressure layers between 100~bar and 10$^{-9}$~bar, evenly spaced in logarithmic space. The spectra are computed between 2.5\,$\mu$m and 5.5\,$\mu$m at a resolution of R=20,000. The model atmosphere assumes constant with height abundances for the chemical species being considered: H$_2$O \citep{Rothman2010}, CH$_4$ \citep{Yurchenko2014}, H$_2$S \citep{azzam2016}, SO$_2$ \citep{underwood2016}, HCN \citep{Barber2014}, CO \citep{Rothman2010}, and CO$_2$ \citep{Rothman2010}. The models include H$_2$-H$_2$ and H$_2$-He collision induced absorption \citep{Richard2012}.

The vertical temperature structure is parametrized following the prescription from \cite{madhu2009}, although we use an anchor point at the top of the atmosphere for the reference temperature ($10^{-9}$\,bar) rather than at 10~mbar used in the \textsc{Poseidon} analysis \citep[see][for a discussion of potential differences that can arise from the anchor point choice]{Welbanks2022}. 
% While these two treatments of the vertical temperature structure are meant to be interchangeable, their differences as well as those associated from the different prior-spaces can result in different retrieved chemical abundances \citep[see e.g.,][]{Welbanks2022}.
The presence of aerosols is included by following the inhomogeneous cloud treatment from \citet{Line2016}, using a cloud-free sector and a sector with a gray cloud deck at a given atmospheric pressure and deviations from H$_2$ Rayleigh-scattering using the prescription from \cite{Lecavelier2008a} as described in \cite{Welbanks2021}. The model atmosphere has both reference points P$_{\rm{ref}}$ and R$_{\rm{P}}$ as free parameters \citep[see e.g., ][]{Welbanks2019a}.

The analysis with Aurora includes three parameters for the effects of an inhomogeneous stellar surface and a parameter for instrumental offsets. First, we retrieve for the photospheric temperature of the star through T$_{\rm{phot}}$, then inhomogeneities at a given temperature T$_{\rm{het}}$ are included by considering that they cover a fraction $f_{\rm{het}}$ of the photosphere during transit. We also consider the possibility of offsets between both NIRSpec detectors using NRS1 as the reference point.

We adopt the same priors as \textsc{Poseidon} as presented in Table~\ref{tab:retrieval_priors}, but since we place the top of the atmosphere at 10$^{-9}$~bar for Aurora (instead of 10$^{-7}$~bar) we use this for the lowest pressure in the priors for the pressure-related parameters. We tested two treatments for the prior of $T_0$ (the top of atmosphere temperature); we first adopt the same prior range as adopted in \textsc{Poseidon} for $T_{\rm{ref}}$ and second we adopt a lower limit of 800~K following the approach in \cite{Welbanks2019b}. The later prior represents the preferred solution for Aurora as it does not allow for unphysically low temperatures at the top of the atmosphere.

\subsection{Retrieval results} \label{subsec:retrieval_results}

\begin{table*}
    \centering
    \caption{Retrieval results for WASP-52\,b's \textit{JWST} NIRSpec/G395H transmission spectrum. For the non-detected chemical species, we present their 2$\sigma$ upper limits. Detailed posterior distribution is shown in Figure \ref{fig:reductions}.}
    \label{tab:retrieval_results}
    \small
    \renewcommand{\arraystretch}{1.2}
    \begin{tabular}{@{} >{\raggedright\arraybackslash}p{3.1cm} *{7}{>{\centering\arraybackslash}p{1.8cm}} @{}}
        \hline\hline
        Reduction Pipeline & \multicolumn{3}{c}{\texttt{transitspectroscopy}} & \multicolumn{2}{c}{\texttt{FIREFly}} & \multicolumn{2}{c}{\texttt{Default JWST}} \\
        Retrieval Code & \textsc{POSEIDON} & \textsc{POSEIDON} & \textsc{Aurora} & \textsc{POSEIDON} & \textsc{POSEIDON} & \textsc{POSEIDON} & \textsc{POSEIDON} \\
        \cmidrule(lr){2-4} \cmidrule(lr){5-6} \cmidrule(lr){7-8}

        Parameter & Atmosphere Only & Atmosphere + Spots & Atmosphere + Spots & Atmosphere Only & Atmosphere + Spots & Atmosphere Only & Atmosphere + Spots \\
        \hline
        \textbf{P-T Profile} & & & & & \\
        $\rm{T_{ref}}$ (K) & $1021.4^{+134.8}_{-126.7}$ & $774.4^{+138.3}_{-118.2}$ & $849.1^{+59.4}_{-34.0}$ & $1137.0^{+116.8}_{-118.2}$ & $892.1^{+143.3}_{-115.0}$ & $835.9^{+124.0}_{-90.1}$ & $835.1^{+114.7}_{-85.3}$ \\
        $\alpha_1$  & $1.12^{+0.53}_{-0.54}$ & $1.11^{+0.55}_{-0.56}$ & $1.62^{+0.25}_{-0.32}$ & $1.09^{+0.57}_{-0.56}$ & $1.08^{+0.57}_{-0.57}$ & $1.12^{+0.55}_{-0.55}$ & $1.07^{+0.55}_{-0.52}$ \\
        $\alpha_2$  & $1.15^{+0.52}_{-0.53}$ & $1.13^{+0.55}_{-0.57}$ & $1.26^{+0.46}_{-0.55}$ & $1.15^{+0.53}_{-0.54}$ & $1.08^{+0.56}_{-0.55}$ & $1.12^{+0.53}_{-0.52}$ & $1.09^{+0.55}_{-0.54}$ \\
        $\log(P_1 / \text{bar})$  & $-3.71^{+2.35}_{-2.03}$ & $-3.52^{+2.43}_{-2.18}$ & $-2.29^{+1.85}_{-2.31}$ & $-3.64^{+2.32}_{-2.12}$ & $-3.35^{+2.35}_{-2.24}$ & $-3.56^{+2.28}_{-2.07}$ & $-3.28^{+2.20}_{-2.26}$ \\
        $\log(P_2 / \text{bar})$  & $-3.30^{+2.22}_{-2.27}$ & $-3.43^{+2.32}_{-2.22}$ & $-5.83^{+2.27}_{-1.95}$ & $-3.33^{+2.21}_{-2.19}$ & $-3.47^{+2.18}_{-2.20}$ & $-3.59^{+2.25}_{-2.07}$ & $-3.49^{+2.22}_{-2.17}$ \\
        $\log(P_3 / \text{bar})$  & $0.45^{+1.03}_{-1.32}$ & $0.46^{+1.04}_{-1.32}$ & $0.49^{+0.98}_{-1.18}$ & $0.39^{+1.06}_{-1.28}$ & $0.45^{+1.02}_{-1.30}$ & $0.33^{+1.07}_{-1.29}$ & $0.41^{+1.02}_{-1.21}$ \\
        \hline
        
        \textbf{Composition} & & & & & \\
        log H$_2$O & $-4.53^{+1.31}_{-0.62}$ & $-3.41^{+1.04}_{-1.04}$ & $-3.03^{+1.21}_{-1.18}$ & $-4.91^{+1.31}_{-0.54}$ & $-3.64^{+1.19}_{-1.21}$ & $-2.39^{+0.66}_{-1.42}$ & $-2.42^{+0.67}_{-1.18}$ \\
        log CO$_2$ & $-7.11^{+1.11}_{-0.50}$ & $-5.09^{+1.09}_{-1.14}$ & $-5.71^{+0.84}_{-0.93}$ & $-7.43^{+1.11}_{-0.43}$ & $-5.67^{+1.18}_{-1.23}$ & $-5.70^{+0.65}_{-1.24}$ & $-5.70^{+0.66}_{-1.04}$ \\
        log H$_2$S & $-8.44^{+2.25}_{-2.21}$ & $-4.24^{+0.79}_{-1.20}$ & $-4.21^{+0.87}_{-0.87}$ & $-7.26^{+1.70}_{-2.84}$ & $-4.44^{+0.93}_{-1.02}$ & $-3.87^{+0.67}_{-1.26}$ & $-3.84^{+0.64}_{-0.97}$ \\
        % \textcolor{red}{log CO} & $-7.69^{+1.78}_{-2.58}$ & $-4.50^{+1.34}_{-2.45}$ & $-8.45^{+1.87}_{-2.15}$ & $-5.12^{+1.68}_{-2.76}$ & $-5.73^{+2.05}_{-3.33}$ & $-5.66^{+2.18}_{-3.50}$ \\
        log CO & $<-4.13$ & $<-1.82$ & $<-2.67$ & $<-4.71$ & $<-1.76$ & $<-1.63$ & $<-1.30$ \\
        log CH$_4$ & $<-8.06$ & $<-7.39$ & $<-7.30$ & $<-8.39$ & $<-7.84$ & $<-7.33$ & $<-7.27$ \\
        log HCN & $<-5.11$ & $<-4.03$ & $<-4.86$ & $<-6.34$ & $<-5.74$ & $<-4.12$ & $<-4.21$ \\
        log SO$_2$ & $<-7.02$ & $<-6.61$ & $<-6.58$ & $<-7.05$ & $<-6.53$ & $<-6.40$ & $<-6.45$ \\
        log K & $<-1.45$ & $<-0.98$ & --- & $<-1.51$ & $<-0.90$ & $<-1.19$ & $<-0.93$ \\
        \hline
        
        \textbf{Aerosols} & & & & & \\
        log $a$ & $3.35^{+3.57}_{-4.71}$ & $1.41^{+3.83}_{-3.39}$ & $2.88^{+4.38}_{-4.26}$ & $2.04^{+3.97}_{-3.85}$ & $1.45^{+3.78}_{-3.42}$ & $1.62^{+3.79}_{-3.46}$ & $1.76^{+3.66}_{-3.47}$ \\
        $\gamma$ & $-7.29^{+2.74}_{-7.69}$ & $-9.94^{+6.25}_{-6.10}$ & $-9.30^{+5.81}_{-6.50}$ & $-9.65^{+5.28}_{-6.46}$ & $-9.77^{+6.34}_{-6.32}$ & $-10.57^{+6.05}_{-5.76}$ & $-10.63^{+6.30}_{-5.68}$ \\
        log P$_{\rm cloud}$ (bar) & $-1.55^{+1.81}_{-1.11}$ & $-3.80^{+1.13}_{-1.11}$ & $-3.74^{+2.60}_{-2.69}$ & $-1.52^{+1.81}_{-1.25}$ & $-3.88^{+1.32}_{-1.18}$ & $-1.22^{+1.95}_{-1.87}$ & $-1.35^{+2.00}_{-1.93}$ \\
        $\phi$ & $0.55^{+0.30}_{-0.25}$ & $0.45^{+0.18}_{-0.15}$ & $0.28^{+0.23}_{-0.13}$ & $0.51^{+0.32}_{-0.30}$ & $0.37^{+0.16}_{-0.14}$ & $0.39^{+0.35}_{-0.26}$ & $0.37^{+0.35}_{-0.23}$ \\
        \hline
        
        \textbf{Stellar} & & & & & \\
        $T_{\rm phot}$ (K) & --- & $5013.1^{+88.3}_{-89.6}$ & $5007.4^{+85.4}_{-85.4}$ & --- & $5006.0^{+90.2}_{-93.3}$ & --- & $5013.0^{+84.8}_{-85.2}$ \\
        $T_{\rm spot}$ (K) & --- & $3736.4^{+184.6}_{-156.4}$ & $3825.4^{+1265.8}_{-211.6}$ & --- & $3804.0^{+188.2}_{-183.5}$ & --- & $4953.0^{+354.5}_{-321.2}$ \\
        $f_{\rm spot}$ & --- & $0.14^{+0.03}_{-0.03}$ & $0.09^{+0.04}_{-0.03}$ & --- & $0.11^{+0.04}_{-0.03}$ & --- & $0.15^{+0.13}_{-0.09}$ \\

        \hline
        
        \textbf{Other Parameters} & & & & & \\
        ${\rm R_{p, ref}}$ (R$_{\rm Jup}$) & $1.23^{+0.01}_{-0.01}$ & $1.20^{+0.01}_{-0.01}$ & $1.31^{+0.03}_{-0.03}$ & $1.23^{+0.01}_{-0.01}$ & $1.20^{+0.01}_{-0.01}$ & $1.24^{+0.01}_{-0.01}$ & $1.23^{+0.01}_{-0.01}$ \\
        $\delta_{\rm rel}$ (ppm) & $-307.99^{+98.07}_{-52.28}$ & $-187.02^{+47.68}_{-47.38}$ & $-196.98^{+50.04}_{-49.85}$ & $-174.41^{+51.07}_{-49.60}$ & $-79.90^{+45.24}_{-44.76}$ & $-165.21^{+53.02}_{-55.61}$ & $-170.31^{+54.85}_{-53.16}$ \\
        \hline
        
        \textbf{Statistics} & & & & &  \\
        Data points & 156 & 156 & 156  & 156  & 156  & 165  & 165  \\
        Degrees of freedom & 136  & 133  & 133  & 136  & 133  & 145  & 142   \\
        ${\chi}^2_\nu$ & 1.18 & 1.16 & 1.20 & 1.07 & 1.05 & 1.06 & 1.07 \\
        $\ln \mathcal{Z}_{\mathrm{Bayesian}}$ & 1113.18 & 1118.83 & 1113.69 & 1122.95 & 1126.10 & 1135.07 & 1135.05 \\
        $\mathcal{B}_{01}$ & Ref & 286 & N/A & Ref & 232 & Ref & 0.983 \\
        Detection significance for star-spots & Ref & $3.8 \sigma$ & N/A & Ref & $3.0 \sigma$ & Ref & N/A \\
        \hline
        \hline
    \end{tabular}
\end{table*}

%-----------
%\begin{enumerate}
%    \item \textbf{Sensitivity to model configurations plot} - spectra fit mosaic
%    \item \textbf{Sensitivity to reductions plot (POSEIDON)} - spectra fit mosaic
%    \item Sensitivity to retrieval codes plot - spectra fit mosaic
%    \item Stacked posterior histograms or posterior overplots
%\end{enumerate}
%-----------
%Prior tables
%Result tables
%-----------

%Main results:
%\begin{enumerate}
%    \item Detection of trace species with abundances: \ce{H2O}, \ce{CO2}
%    \item Evidence of \ce{CO}, \ce{H2S}
%    \item High altitude cloud with fractional coverage
%    \item significant stellar contamination
%\end{enumerate}

% Appendix: one example corner overplot; Zenodo \\

\begin{figure*}
    \centering\includegraphics[width=1.0\linewidth]{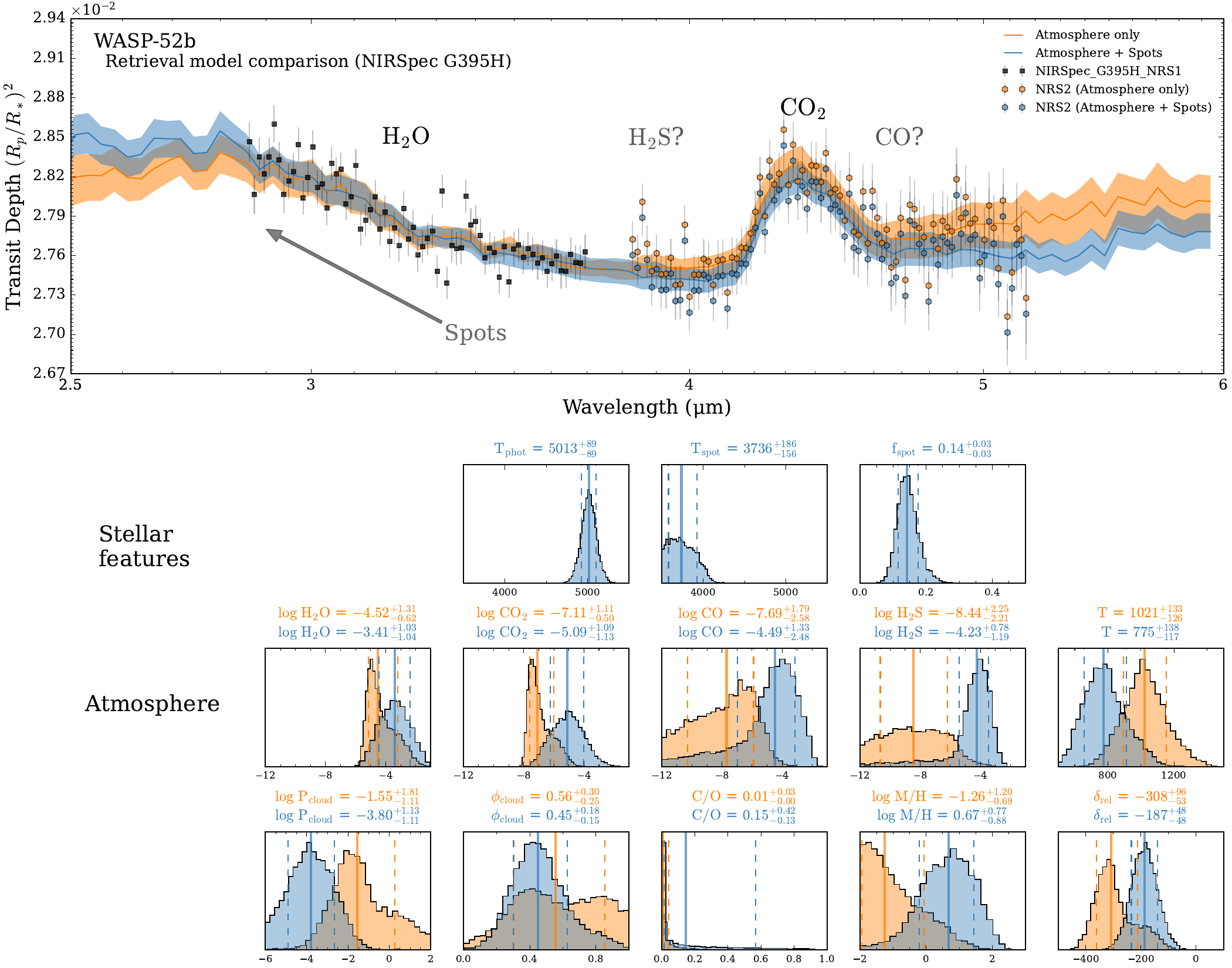}
    \caption{\textsc{POSEIDON} atmospheric and stellar retrieval results for WASP-52b. Top: NIRSpec/G395H transmission spectrum from the \texttt{transitspectroscopy} data reduction, with the two models varying stellar parameters overplotted. The unocculted starspots introduce an upward spectral slope, indicated by the grey arrow. The NIRSpec/G395H NRS2 data are offset by the median retrieved offset from each model. Each model has a median retrieved spectrum (solid line) and the $\pm$ 1$\sigma$ confidence interval (shaded contours). Bottom: retrieval models' posterior distributions with their median values (solid lines) and their $\pm$ 1$\sigma$ confidence intervals (dashed lines). The stellar contamination parameters are highlighted in the top row, and atmospheric parameters are highlighted in the middle and bottom rows (see Table \ref{tab:retrieval_results} for full retrieval results).}
    \label{fig:models}
\end{figure*}

\begin{figure*}
    \centering\includegraphics[width=1.0\linewidth]{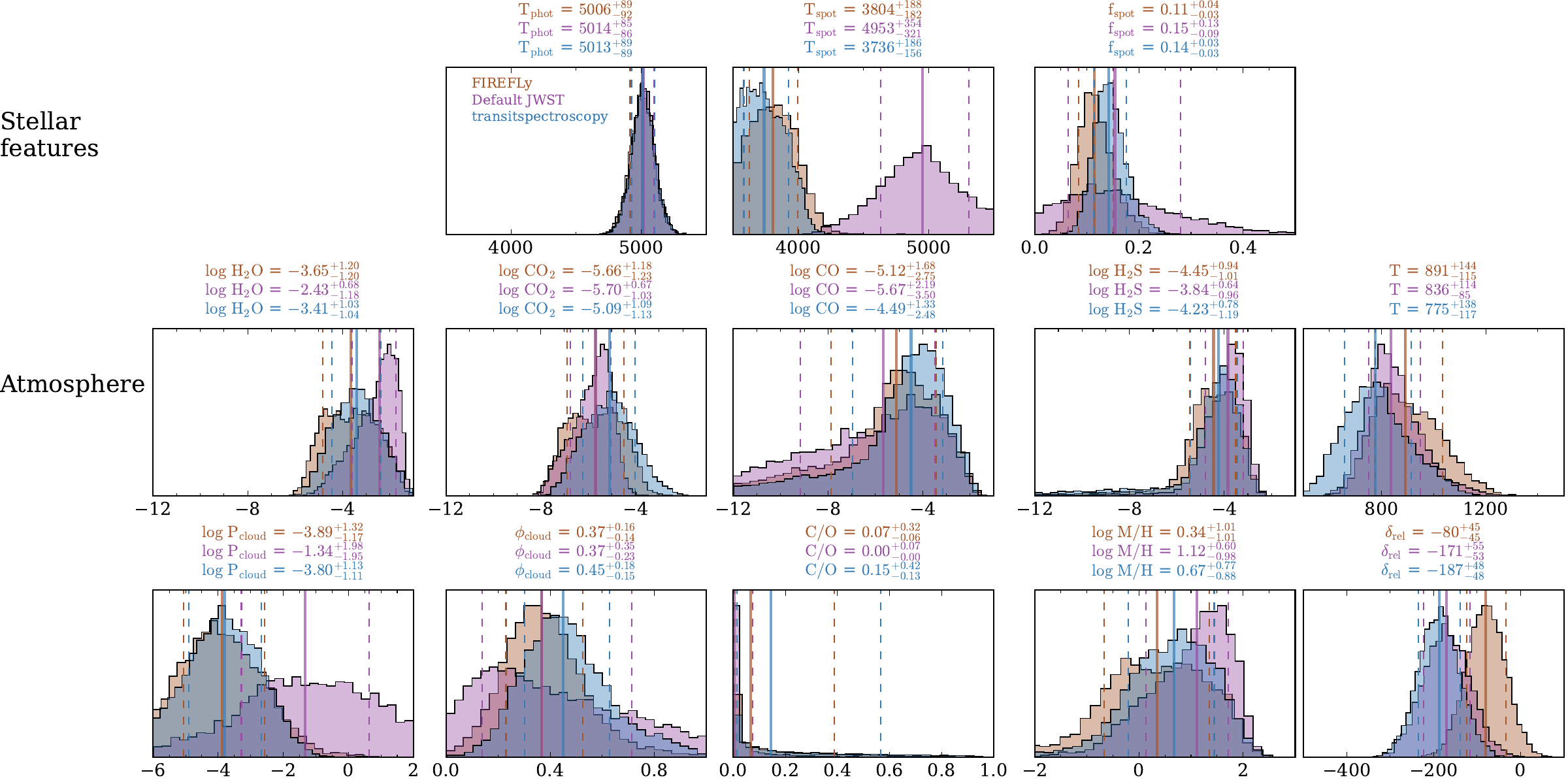}
    \caption{Retrieval posterior distributions for different data reductions with their median values (solid lines) and their $\pm$ 1$\sigma$ confidence intervals (dashed lines). The retrieval results for \texttt{FIREFly} (brown), \texttt{Default JWST} (purple), and \texttt{transitspectroscopy} (blue) are shown for the statistically preferred atmosphere + spots model. Top: retrieved stellar contamination parameters for the atmosphere + spots model. Middle and bottom: key retrieved atmospheric properties from both models (see Table \ref{tab:retrieval_results} for full retrieval results).}
    \label{fig:reductions}
\end{figure*}

\begin{figure*}
    \centering\includegraphics[width=1.0\linewidth]{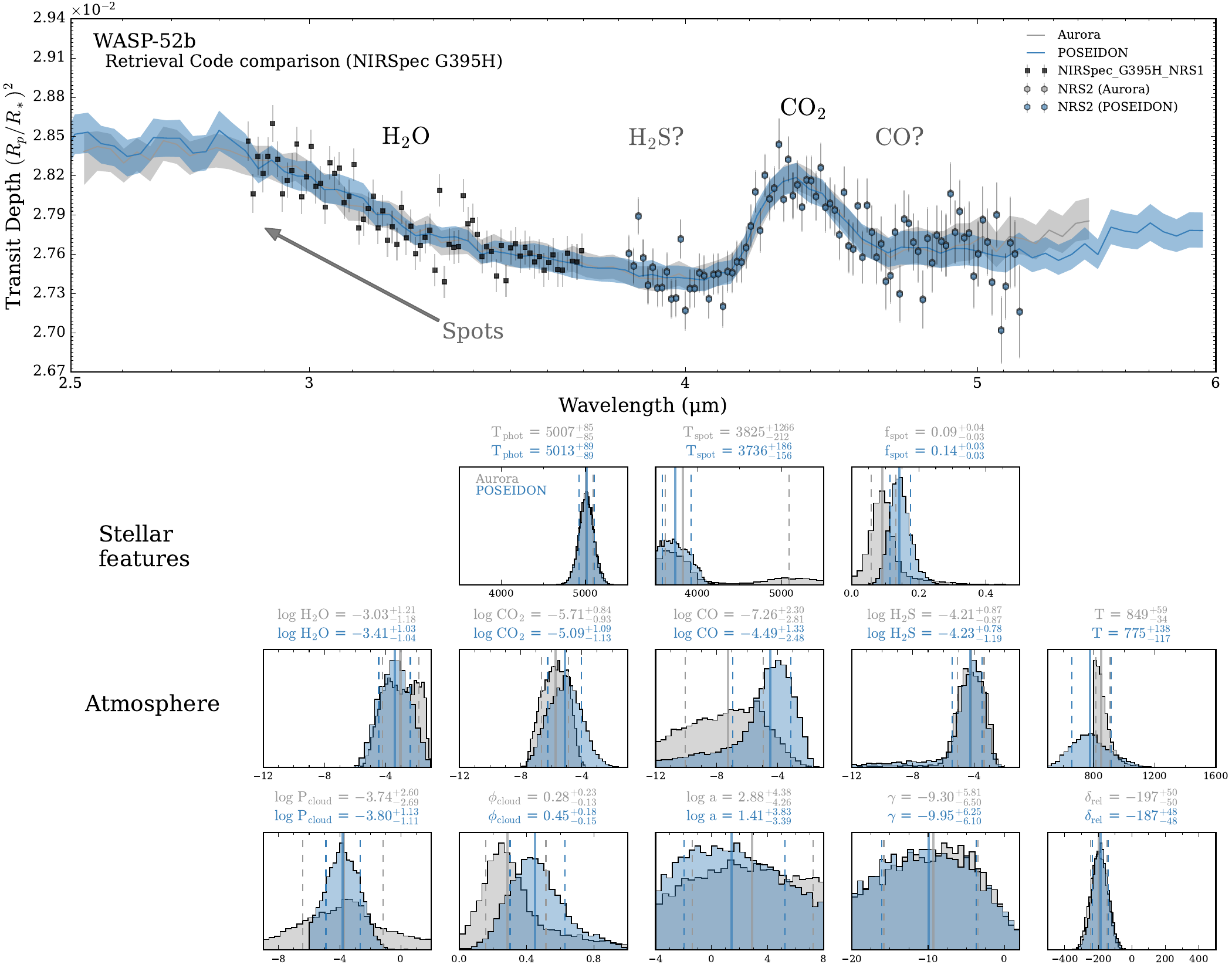}
    \caption{Aurora and POSEIDON retrieval code comparison. Top: NIRSpec/G395H transmission spectrum from the \texttt{transitspectroscopy} reduction, with the two models varying retrieval codes overplotted. Bottom: Retrieval posterior distributions with their median values (solid lines) and their $\pm$ 1$\sigma$ confidence intervals (dashed lines). The POSEIDON (blue) and Aurora (grey) results are shown for the statistically preferred atmosphere + spots model (see Table \ref{tab:retrieval_results} for full retrieval results).}
    \label{fig:aurora}
\end{figure*}

WASP-52b's JWST NIRSpec/G395H transmission spectrum reveals significant \ce{H2O} and \ce{CO2} detections alongside evidence for \ce{H2S}, with no evidence for other chemical species. In addition, we find unocculted starspots on its host star are needed to explain WASP-52b's NIRSpec/G395H spectrum ($3.8 \sigma$ for \texttt{transitspectroscopy} and $3.0 \sigma$ for \texttt{FIREFLy}). Regardless of which stellar contamination model is used, our results suggest high-altitude clouds or hazes (with partial coverage at the terminator), in order to match the observations. In the following sections, we report quantitative details on these atmospheric and stellar inferences in Figure \ref{fig:reductions} and Table \ref{tab:retrieval_results} and compare our \textsc{POSEIDON} and Aurora results in Figure~\ref{fig:aurora}. We find consistent retrieval results between the \texttt{transitspectroscopy} and \texttt{FIREFLy} pipelines, but we note that several differences arise in the results from the default JWST pipeline (e.g. in the retrieval starspot properties, see Section~\ref{subsubsec:starspots} and Figure~\ref{fig:reductions}). These likely arise from the greater statistical scatter in this reduction, with several peaks  in the NRS2 data not seen for \texttt{transitspectroscopy} or \texttt{FIREFLy} (see Figure~\ref{fig:obs}, so we consider this reduction lower-quality for the purposes of drawing atmospheric and stellar constraints. We therefore adopt the \texttt{transitspectroscopy} as our default reduction pipeline and present its atmospheric retrieval results in Section \ref{subsubsec:atmosphere}.

\subsubsection{The atmosphere of WASP-52b} \label{subsubsec:atmosphere}
Our retrieval results\footnote{Atmosphere+Spots retrieval on \texttt{transitspectroscopy}} indicate a chemically rich atmosphere on WASP-52b, showing significant detections of \ce{H2O} and \ce{CO2}, and evidence for \ce{H2S}. We detect \ce{H2O} with high significance (log $\rm H_2 O=-3.41^{+1.04}_{-1.04}; \: \mathcal{B}_{01}:370 \: [3.9 \sigma$]), with the absorption feature tailing the 2.8-3.5 $\mu$m range. The \ce{H2O} feature is potentially obscured by atmospheric hazes, which have a similar upward slope effect on the spectrum. For the first time for WASP-52b, the \ce{CO2} feature near 4.3 $\mu$m is detected (log $\rm C O_2=-5.09^{+1.09}_{-1.14}; \: \mathcal{B}_{01}:3.6e23 \: [>10 \sigma$]). These retrieved abundances do not strongly constrain metallicity and C/O ratio (C/O=$0.07^{+0.32}_{-0.06}$).

Moreover, we find evidence for \ce{H2S} (log $\rm H_2S=-4.24^{+0.79}_{-1.20}; \: \mathcal{B}_{01}:3.11 \: [2.1 \sigma]$). The \ce{H2S} evidence is still tentative, as its long posterior tails in Figure \ref{fig:models} suggest. Due to the NIRSpec/G395H wavelength range lacking sensitivity to the \ce{K} feature near 0.8 $\mu$m, we could not find useful constraints on the \ce{K} abundances with this observation. There is also no evidence of \ce{CH4}, \ce{NH3}, \ce{HCN}, seen in our WASP-52b NIRSpec G395H spectrum.

The relatively muted spectral features further suggests the presence of high-altitude clouds ($\ln \mathcal{B_{01}} = 1.9$ / $\sim 2.5\,\sigma$ for \texttt{transitspectroscopy}). The cloud deck has a cloud-top pressure of $\log P_{\rm{cloud}} = -3.80^{+1.13}_{-1.11}$ ($\sim$ 0.1\,mbar) and covers $45^{+18}_{-15}$\% of the terminator. The inference of clouds is consistent with the previous HST study that inferred a cloudy atmosphere for WASP-52b \citep{Bruno2018HST}, though our relatively low significance for clouds is driven by the degeneracy between stellar contamination and clouds \citep{fournier_tondreau2024, fournier_tondreau2025}.

We find generally good agreement between \textsc{POSEIDON} and Aurora (Figure~\ref{fig:aurora}). Both codes identify H$_2$O and CO$_2$ as the main absorbers contributing to WASP-52b's transmission spectrum with consistent abundances (log $\rm H_2 O = -3.41^{+1.04}_{-1.04}$ [\textsc{POSEIDON}] vs log $\rm H_2 O = -3.03^{+1.21}_{-1.18}$ [Aurora]; log $\rm CO_2 = -5.09^{+1.09}_{-1.14}$ [\textsc{POSEIDON}] vs log $\rm CO_2 = -5.71^{+0.84}_{-0.93}$ [Aurora]) and find similar tentative evidence for H$_2$S (log $\rm H_2 S=-4.24^{+0.79}_{-1.20}$ [\textsc{POSEIDON}] vs log $\rm H_2 S=-4.21^{+0.87}_{-0.87}$ [Aurora]). Therefore, the small differences in the retrieval configurations (e.g. the different temperature parameter lower limits and the different top-of-atmosphere pressures) do not affect our conclusions for WASP-52b's atmospheric composition.

% while The cloud's patchiness offers us an explanation to this planet's historically variable spectral features \citep{kirk2016, chen2017, Bruno2018HST, bruno2020, Chen2020, fournier_tondreau2025}.} 

\subsubsection{Unocculted starspots on WASP-52} \label{subsubsec:starspots}

Moreover, we have assessed whether unocculted stellar active regions are needed to account for the NIRSpec/G395H spectrum of WASP-52b. The lower sections of Table \ref{tab:retrieval_results} present a summary of fit quality and comparisons between models through metrics such as reduced chi-square ${\chi}^2_\nu$, log Bayesian evidence $\ln \mathcal{Z}_{\mathrm{Bayesian}}$, Bayes factor $\mathcal{B}_{01}$, and detection significances. Retrievals using various stellar contamination models and data reduction pipelines reveal that the atmosphere+starspots model is favored over the atmosphere-only model ($\ln \mathcal{B} =$ 5.7 / $\sim$3.8$\sigma$ for \texttt{transitspectroscopy} and  $\ln \mathcal{B} =$ 3.2 / $\sim$3.0$\sigma$ for \texttt{FIREFly}). Nonetheless, the atmosphere-only model (the yellow model spectrum in figure \ref{fig:models}) gives a reasonable fit to the NIRSpec/G395H data (${\chi}^2_\nu \approx 1.1$), and cannot be easily ruled out. The spectral fit's log Bayesian evidence improvement mainly comes from adding a spectral slope from starspots. While the atmosphere+spots+faculae model has been shown to be the optimal model in the SOSS observations of \cite{fournier_tondreau2025}, 
we find the additional contribution of faculae is not significant at the longer wavelengths of NIRSpec/G395H. Therefore, we have only used the single-heterogeneity (i.e. starspots) in this paper. In this case, we will refer to the atmosphere+spots model as the ``preferred'' model.

The preferred model indicates the presence of unocculted starspots on WASP-52. The \texttt{transitspectroscopy} reduced spectrum indicates starspots covering $14^{+3}_{-3}$ percent of the visible stellar surface, with its temperature $\approx$ 1250K cooler than the photosphere. We find similar results from the \texttt{FIREFly} reduced spectrum, showing $11^{+4}_{-3}$ percent of the visible stellar surface is $\approx$ 1200K cooler than its photosphere. These inferences indicate the presence of unocculted starspots, in agreement with previous HST \citep{Bruno2018HST, bruno2020} and JWST NIRISS studies \citep{fournier_tondreau2025}.

We note that our retrieved stellar properties from the default JWST pipeline differ somewhat from those from \texttt{transitspectroscopy} and \texttt{FIREFly}. The default JWST pipeline spectrum instead favors starspots with a similar temperature to the photosphere temperature (Figure~ \ref{fig:reductions}), which indicates no significant level of stellar contamination. In this case, the lack of stellar contamination is compensated by additional water absorption from the higher H$_2$O abundance (log \ce{H2O} =$-2.42^{+0.67}
_{-1.18}$), since the half of the 3\,$\micron$ H$_2$O absorption feature covered by NIRSpec G395H can compensate for the starspots' upward slope. Therefore, for this reduction only, we did not find a preference for the atmosphere + spots model over the atmosphere-only model. However, the good agreement between \texttt{transitspectroscopy} and \texttt{FIREFly}, combined with the strong existing evidence of stellar contamination on WASP-52 from JWST \citep{fournier_tondreau2025} and Hubble \citep[e.g.][]{bruno2020}, means we favor the results from \texttt{transitspectroscopy} as our default interpretation for stellar contamination in WASP-52b's NIRSpec G395H spectrum.

\section{Discussion \& Conclusion} \label{sec:discu}

In this paper, we presented and analyzed the JWST NIRSpec/G395H limb-averaged transmission spectrum of WASP-52b. Our atmospheric retrievals robustly detect \ce{H2O} and \ce{CO2}, making this the first clear detection of \ce{CO2} in WASP-52b. We also find tentative evidence for \ce{H2S}. The spectrum exhibits an upward slope and partially muted spectral features, which can be attributed to an inhomogeneous aerosol layer with partial terminator coverage and the presence of unocculted starspots. Furthermore, we find that including unocculted starspots is statistically preferred over only having atmospheric parameters in the retrieval models. We proceed to discuss implications of our analysis.

\subsection{Additional Trace Species in \\ WASP-52b's Atmosphere}

The tentative evidence for \ce{H2S} ($\mathcal{B}_{01} \approx 3$) is intriguing, as sulfur-bearing species are tracers for both planet formation processes and photochemistry. The recent  detection of \ce{H2S} in HD 189733b by \cite{fu2024} using JWST NIRCam. illustrates that it is possible to detect this molecule. However, the reliability of the \ce{H2S} inference in our case is complicated by degeneracies between unocculted starspots, inhomogeneous clouds, and the weak \ce{H2S} band near 4\,$\micron$. 

% \cite{feinstein2023} has detected \ce{CO} at 3.6 $\sigma$ significance with signatures for \ce{CO2}, which help to determine the sub-solar C/O ratio in WASP-39b, using JWST NIRISS. Further study by \cite{alderson2023} presented the significant absorption feature from \ce{CO2} to 28.5 $\sigma$ and confirmed WASP-39b can be best described by an atmospheric model with 3-10 times solar metallicity with sub-solar to solar C/O ratios, using JWST NIRSpec G395H. 

We also note that the non-detection of \ce{CO} in our NIRSpec G395H spectrum hinders our ability to constrain the C/O ratio in this analysis. Nevertheless, we observe a suggestive peak in the CO abundance posteriors near log CO $\sim -4$ (e.g. Figure~\ref{fig:reductions}), which suggests that additional observations covering the 4.5--5.2\,$\micron$ region could resolve the contribution of CO to WASP-52b's transmission spectrum and hence constrain its C/O ratio.

\subsection{Evidence of Patchy Hazes}

Our retrieval results suggest WASP-52b to have patchy aerosols. This inhomogeneity offers a natural explanation for the variability of WASP-52b's transmission spectrum at different epochs. The patchy cloud suggests the cloud deck is not uniform across the terminator, potentially due to the pressure and temperature difference between the morning and evening limbs \citep{Line2016, espinoza2021}. Although our current retrieval models assume an atmosphere with a 1D pressure-temperature profile with patchy cloud, the inferred patchiness of the cloud deck can hint at the difference between the morning and evening terminators. With this key result from the JWST MEP, we will also present and analyze the asymmetric spectra of the morning and evening limbs on WASP-52b in subsequent papers.

\subsection{Time-variable Stellar Contamination in WASP-52b Transmission Spectra}

\begin{figure*}
    \centering\includegraphics[width=1.0\linewidth]{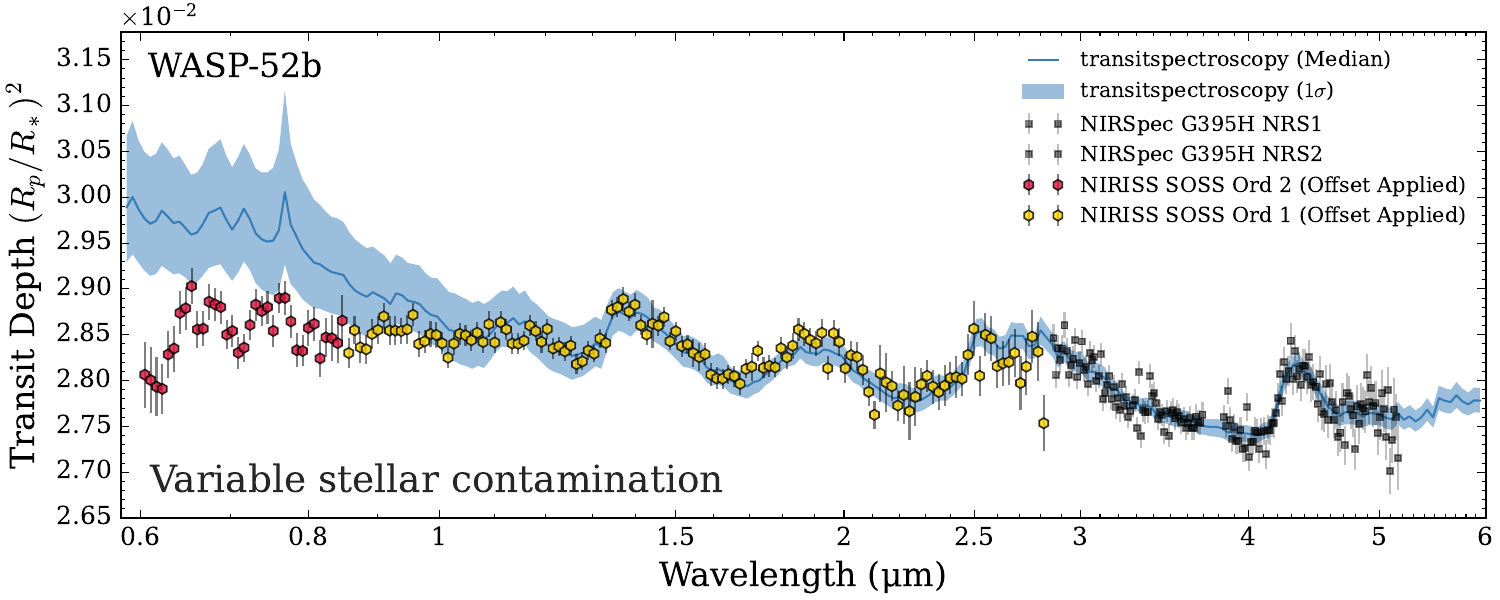}
    \caption{The NIRISS/SOSS \citep{fournier_tondreau2025} and NIRSpec/G395H combined spectra overplotted with the atmosphere+spots spectral model fitted with the NIRSpec/G395H \textsc{transitspectroscopy} data. The varying slope towards the short wavelength ($\le 1.5 \micron$) is potentially indicative of the time variability of stellar active regions (starspots and faculae) on the host star.}
    \label{fig:stellar_contam}
\end{figure*}

A central challenge in analyzing WASP-52b's spectrum is the presence of unocculted stellar contamination on its host star. The spectral slope induced by the unocculted starspots and faculae is degenerate with the absorption tail of the water vapor. This is illustrated by our retrieval results: the \texttt{transitspectroscopy} and \texttt{FIREFly} reduction favor the spots+atmosphere model, inferring a spot coverage fraction and spot temperature consistent with the NIRISS analysis by \cite{fournier_tondreau2025}. In contrast, the spectrum reduced by the default JWST pipeline shows little preference for spots, instead compensating with a higher, potentially biased, \ce{H2O} abundance. Furthermore, the combined NIRISS/SOSS and NIRSpec/G395H spectrum (Figure~\ref{fig:stellar_contam}) shows the stellar contamination slope varies across different epochs and suggests WASP-52's stellar regions are highly active. Due to the time-varying nature of stellar contamination for WASP-52b's transmission spectra, standard atmospheric+stellar retrieval approaches are not suitable for the combined JWST NIRISS/SOSS + NIRSpec/G395H dataset. We verified this by attempting a combined retrieval of the NIRISS/SOSS + NIRSpec/G395H data, which resulted in an unphysically high H$_2$O abundance ($\gtrsim$ 10\%; not shown). We reserve such exploration of time-varying stellar contamination for a future study.

In this case, small differences in the reduced spectrum, as shown in Figure \ref{fig:obs}, combined with time-dependent stellar variability, can yield statistically significant differences in the inferred stellar properties. Therefore, this analysis highlights the importance of data reduction comparison and accounting for temporal stellar variability in disentangling stellar and planetary signals, especially for active stars like WASP-52. 

\subsection{Future Directions in the JWST Mornings \& Evenings Program for WASP-52b}  \label{subsec:future}

% Our JWST NIRSpec/G395H limb-averaged transmission spectrum of WASP-52b provides the first detection of \ce{CO2} in this hot Saturn's atmosphere, alongside confirmed \ce{H2O} absorption and evidence for \ce{H2S}. This rich spectrum mixes atmospheric features and stellar contamination imprints, with the spectral slope and muted features best explained by a combination of patchy aerosols covering $\approx 40 \pm 20$\% of terminator and unocculted starspots on the active K-dwarf host. These promising results show that chemical abundances and aerosol properties can be reliably constrained for planets orbiting active stars, despite significant stellar contamination. However, this require a careful approach using multiple data reductions and retrieval codes for cross-validation. 

While our tentative inference of inhomogenous aerosols in WASP-52b's atmosphere is promising, in a future study we will analyze the separate morning and evening spectra to directly probe the differences between WASP-52b's terminators. These analyses are uniquely complicated by the prevalence of stellar contamination in this system, which also varies with time. Our subsequent morning and evening analysis will also consider techniques to account for the time-varying stellar contamination to allow a joint retrieval analysis of the NIRISS/SOSS and NIRSpec/G395H spectra (Figure~\ref{fig:stellar_contam}). A follow-up theory paper, currently in preparation, will use GCMs and microphysical cloud models to explore forward models of clouds in WASP-52b's atmosphere. It will specifically test whether limb asymmetries can explain the patchy cloud fraction retrieved here. WASP-52b remains a promising target for probing GCM predictions, as this limb-averaged spectrum analysis and future MEP studies continue to reveal its atmosphere from the shadow of its spotty star.

% WASP-52b remains a promising target to probe predictions of GCMs, which cannot forever hide under the shadow of its spotty star.

% In conclusion, the JWST NIRSpec/G395H  observation reveals WASP-52b to be a distant world with a diverse chemical inventory, orbiting an active star, and potentially masked in patchy clouds. 

% Add how the the transity white-light transit allows for morning and evening analysis

\begin{acknowledgments}

This work is based on observations made with the NASA/ESA/CSA JWST. The data were obtained from the Mikulski Archive for Space Telescopes at the Space Telescope Science Institute, which is operated by the Association of Universities for Research in Astronomy, Inc., under NASA contract NAS 5-03127 for JWST. These observations are associated with program JWST-PID-3969. Support for program JWST-PID-3969 was provided by NASA through a grant from the Space Telescope Science Institute. This research has made use of NASA's Astrophysics Data System Bibliographic Services. This research was supported in part through computational resources and services provided by Advanced Research Computing at the University of Michigan, Ann Arbor. YP acknowledges support from the University of Michigan's Astronomy \& Astrophysics undergraduate program and unwavering guidance from RJM. RJM acknowledges support from NASA through the NASA Hubble Fellowship grant HST-HF2-51513.001, awarded by the STScI, which is operated by the Association of Universities for Research in Astronomy, Inc., for NASA, under contract NAS 5-26555. NM acknowledges support through a UKRI Future Leaders Fellowship [grant number MR/T040866/1], a Science and Technology Facilities Council Consolidated Grant [ST/R000395/1], and a Science and Technology Facilities Council astronomy observation and theory small award [ST/Y00261X/1]. DAC received support from the Max Planck Society.
D.S. acknowledges funding as part of JWST GO program 3969 (PIs: Espinoza, Powell).
BP acknowledges financial support from the Walter Gyllenberg Foundation. M.S. was supported by the Heising-Simons Foundation through a 51 Pegasi b fellowship.
\end{acknowledgments}

% \begin{contribution}
% ...
% \end{contribution}

\section*{Data availability}
In this paper, the JWST data presented were obtained from the Mikulski Archive for Space Telescopes (MAST) at the Space Telescope Science Institute (STScI). The transit observations analyzed can be accessed through \dataset[doi: 10.17909/2zfw-ep42]{https://doi.org/10.17909/2zfw-ep42}.

Data products are available on Zenodo, doi: 10.5281/zenodo.18734230

\facilities{JWST (NIRSpec/G395H) }

\software{\textsc{astropy} \citep{2013A&A...558A..33A,2018AJ....156..123A,2022ApJ...935..167A}, \textsc{Poseidon} \citep{macDonald2017,macDonald2023}, \textsc{Aurora} \citep{Welbanks2021}, \texttt{transitspectroscopy} \citep{Espinoza:2025}, \texttt{FIREFly} \citep{Rustamkulov2023}, \texttt{FASTCHEM} 2 \citep{Stock2022}, \textsc{MultiNest} \citep{feroz2009}, \textsc{PyMultiNest} \citep{Buchner2014}, \textsc{pandas} \citep{reback2020pandas}, \textsc{numpy} \citep{harris2020array}, \textsc{SciPy} \citep{2020SciPy-NMeth}, \textsc{matplotlib} \citep{Hunter:2007matplotlib}, \textsc{Numba} \citep{lam2015numba}, \textsc{mpi4py} \citep{mpi4py}, \textsc{Conda} \citep{anaconda2016}, \textsc{Jupyter} \citep{Kluyver2016}, \textsc{IPython} \citep{PER-GRA:2007}, \textsc{Python} \citep{van2014python}
}

% \textsc{BeAR} \citep{}, \textsc{Aurora} \citep{}

\appendix

\section{Chemical Equilibrium Retrieval}

Besides our free chemistry retrievals, we also explored retrievals assuming thermochemical equilibrium using \textsc{POSEIDON}. The chemical equilibrium grid uses pre-computed altitude-dependent mixing ratios from \texttt{FASTCHEM} 2 \citep{Stock2022}. An equilibrium retrieval replaces the volume mixing ratios with the atmospheric carbon-to-oxygen ratio and log-metallicity as free parameters. We set the priors for C/O and log$_{10}$ M/H (in solar units) to be $\mathcal{U}(0.2, 1.5)$ and $\mathcal{U}(-1, 4)$, respectively, with the other priors being the same as in Section~\ref{subsec:poseidon_methods}. 

\begin{figure*}[htbp]
    \centering\includegraphics[width=1.0\linewidth]{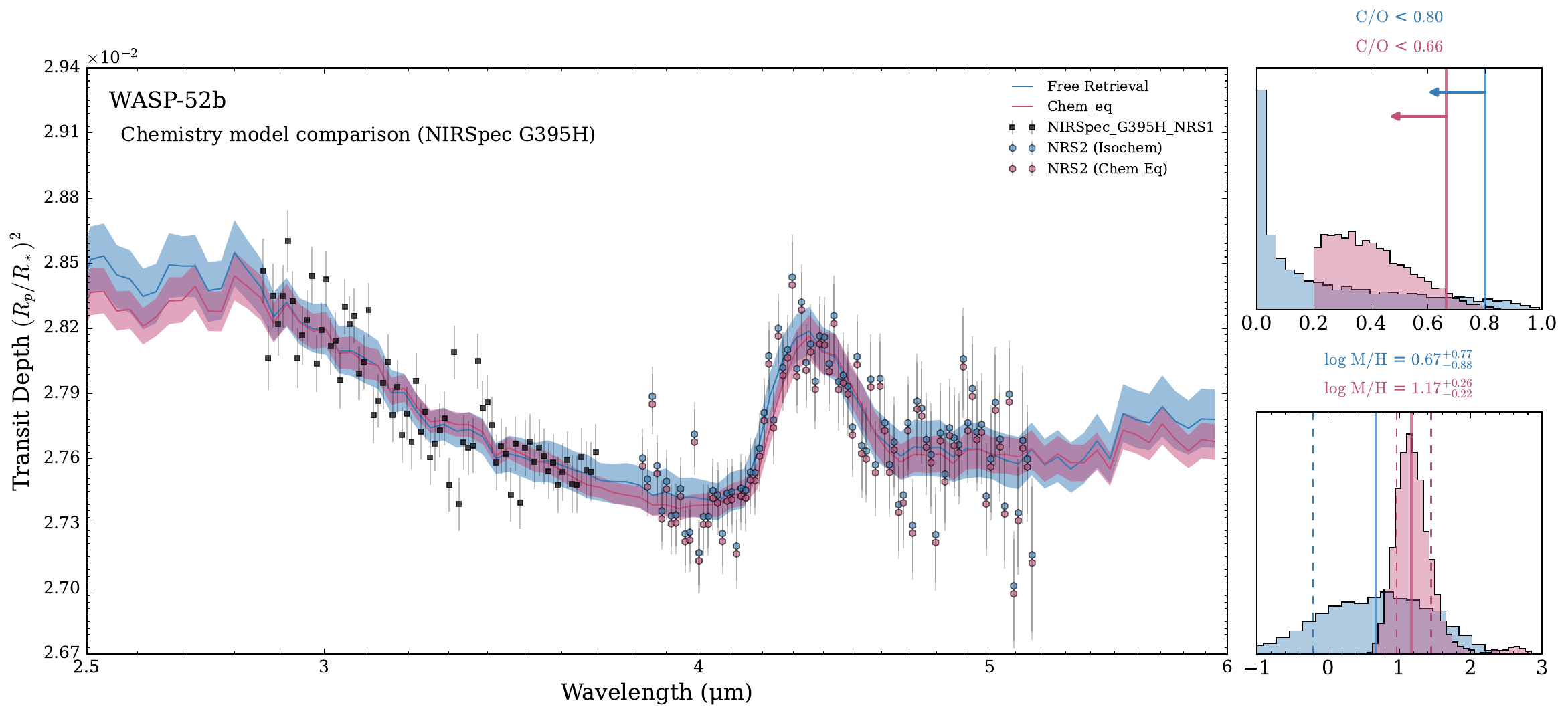}
    \caption{Free and chemical equilibrium retrieval model comparison, similar to Figure \ref{fig:models} and \ref{fig:aurora}. Left: The retrieved spectra by models with different atmospheric chemistry treatments (free retrieval with constant-in-altitude mixing ratios in blue and chemical equilibrium in pink). Right: The retrieved distribution for C/O and metallicity.}
    \label{fig:chem_eq}
\end{figure*}

We compare the chemical equilibrium retrieval with our free retrieval results (using constant-in-altitude mixing ratios) in Figure~\ref{fig:chem_eq}. The chemical equilibrium retrievals find a supersolar metallicity around 15 $\times$ solar ([M/H] $= 1.17^{+0.26}_{-0.22}$, or M/H $= 14.8^{+12.1}_{-5.9} \times$ solar), with a 3$\sigma$ lower limit of 4.3 $\times$ solar and a 5$\sigma$ lower limit of 2.5 $\times$ solar. These tighter constraints compared to the free retrieval are driven by the strong assumption of chemical equilibrium. We also find only an upper limit on the C/O (though note that the lowest C/O ratio in the equilibrium grid is 0.2).

\section{JWST Data Sensitivity to Spitzer-derived Orbital Parameters} \label{appendix:Spitzer}

As mentioned in Section \ref{subsec:spectra}, we repeated the \texttt{transitspectroscopy} spectrum with the previous Spitzer orbital parameters \citep{alam2018}. To be more specific, we changed the period from 1.74978117 days to 1.749779800 days, changed the impact paramter from 0.6007 to 0.6079, and changed a/Rstar from 7.219 to 7.22.

% \textcolor{red}{Add new data comparison figure here @Yanbo @Nestor}

\begin{figure}
    \centering
    \includegraphics[width=1\linewidth]{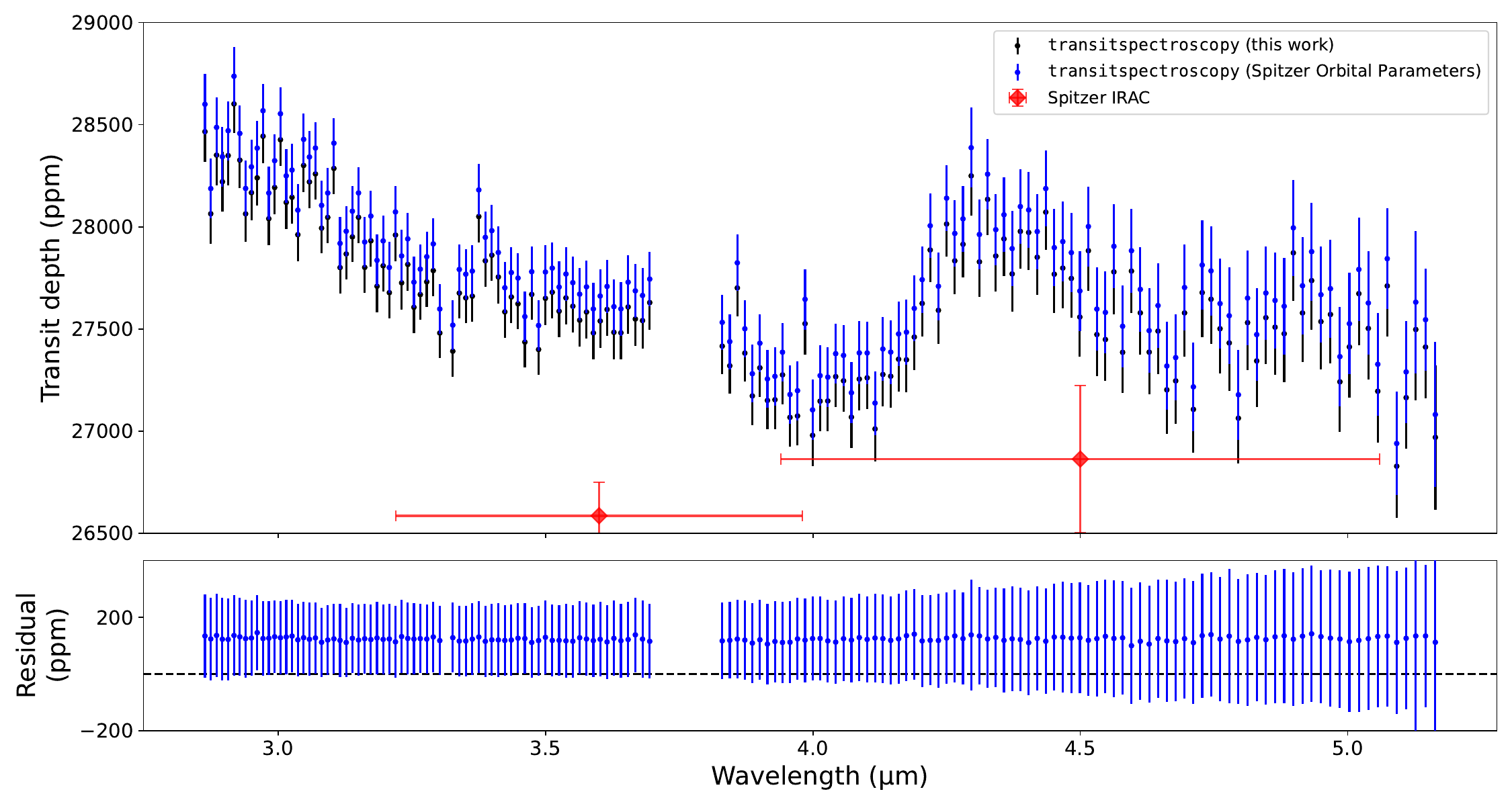}
    \caption{Sensitivity of our WASP-52b NIRSpec/G395H spectra to orbital parameters. Top: \texttt{transitspectroscopy} spectra using orbital parameters derived from the JWST NIRSpec/G395H data (black data) and with orbital parameters fixed to the \emph{Spitzer} analysis from \citet{alam2018} (blue data). The \emph{Spitzer} data are overlaid for comparison (red data). Bottom: difference between the NIRSpec/G395H spectra for the two orbital parameter treatments, showing a uniform $\sim$124 ppm offset.}
    \label{fig:Spitzer_fit}
\end{figure}

% \section{Additional Materials}

% \section{Additional Reduction} \label{sec:pipeline}

\bibliography{ref}{}
\bibliographystyle{aasjournalv7}

\end{CJK*}
\end{document}